\documentclass[11pt,a4paper]{article}

\usepackage[utf8]{inputenc}
\usepackage[T1]{fontenc}
\usepackage{amsmath,amssymb}
\usepackage{graphicx}
\usepackage{booktabs}
\usepackage{hyperref}
\usepackage{url}
\usepackage{natbib}
\usepackage{authblk}
\usepackage[margin=2.5cm]{geometry}
\usepackage{caption}
\usepackage{subcaption}
\usepackage{float}
\usepackage{enumitem}
\usepackage{xcolor}
\usepackage{lineno}

\hypersetup{
  colorlinks=true,
  linkcolor=blue,
  citecolor=blue,
  urlcolor=blue
}

\title{\textbf{INTERACT: An AI-Driven Extended Reality Framework for Accessible Communication Featuring Real-Time Sign Language Interpretation and Emotion Recognition}}

\author[1]{Nikolaos~D.~Tantaroudas\thanks{Corresponding author: \texttt{nikolaos.tantaroudas@iccs.gr}}}
\author[2]{Andrew~J.~McCracken}
\author[3]{Ilias~Karachalios}
\author[4]{Evangelos~Papatheou}

\affil[1]{Institute of Communications and Computer Systems (ICCS), Iroon Polytechneiou 9, 15773 Zografou, Athens, Greece}
\affil[2]{DASKALOS-APPS, 183 Rue de l'Abb\'{e} Griffon, 01960 P\'{e}ronnas, France}
\affil[3]{National Technical University of Athens, Leof.\ Alimou, Katechaki, Zografou, 15772 Athens, Greece}
\affil[4]{Exeter Small-Scale Robotics Laboratory, Engineering Department, University of Exeter, Exeter EX4~4QF, UK}

\date{}

\begin{document}

\maketitle

\begin{abstract}
Video conferencing has become a cornerstone of professional collaboration; however, the majority of existing platforms provide inadequate support for deaf, hard-of-hearing, and multilingual users. The World Health Organization estimates that more than 430~million individuals globally need rehabilitation services for disabling hearing loss, with projections indicating this number will surpass 700~million by 2050. Conventional accessibility measures are constrained by prohibitive costs, scarce availability, and logistical difficulties. Extended Reality (XR) technologies present novel avenues for building immersive and inclusive communication spaces. This paper introduces INTERACT (Inclusive Networking for Translation and Embodied Real-Time Augmented Communication Tool), an AI-driven XR platform that unifies real-time speech-to-text conversion, International Sign Language (ISL) rendering through 3D avatars, multilingual capability, and emotion recognition within immersive virtual settings. Constructed upon the CORTEX2 framework and deployed on Meta Quest~3 headsets, the platform harnesses cutting-edge AI models, including Whisper for speech recognition, NLLB for multilingual translation, RoBERTa for emotion classification, and Google MediaPipe for gesture extraction. Pilot evaluations were carried out in two stages: one involving technical specialists from academic and industrial sectors and another engaging deaf community members directly. The pilot evaluations yielded 92\% user satisfaction, transcription accuracy surpassing 85\%, and 90\% emotion detection precision. Participants gave a mean overall experience score of 4.6 out of 5.0, with 90\% indicating willingness to engage in subsequent testing rounds. The platform reliably accommodates up to 1{,}000 simultaneous users with negligible latency increase. Stress testing confirmed robust performance under high-demand conditions, with 10{,}000 concurrent requests processed at 900--1{,}000 requests per second with zero failures. This work constitutes the first XR-based video conferencing platform that merges real-time ISL rendering via 3D avatars with multilingual speech conversion and emotion-aware feedback within a unified, immersive solution. The findings underscore considerable potential for reshaping accessibility across educational, cultural, and professional domains. An extended version of this work, including comprehensive pilot data and detailed implementation, has been published as an Open Research Europe article~\citep{tantaroudas2026interact}.
\end{abstract}


\section{Introduction}
\label{sec:introduction}

The accelerated transition towards remote and hybrid working models has rendered video conferencing indispensable for professional, educational, and social exchanges worldwide. Nevertheless, mainstream video conferencing solutions frequently fall short of addressing the varied requirements of deaf, hard-of-hearing, and multilingual participants, thereby erecting considerable obstacles to their full engagement and contribution~\citep{alford2023window}. The World Health Organization reports that over 430~million people across the globe need rehabilitation for disabling hearing loss, a number expected to exceed 700~million by 2050~\citep{who2021hearing}. Concurrently, linguistic barriers within multilingual teams impede effective communication and cooperative outcomes~\citep{franceschini2020removing}. This accessibility deficit is especially pronounced in professional environments where subtle communication, encompassing emotional tone and non-verbal signals, is vital for productive collaboration and decision-making.

Conventional accessibility provisions, including human interpreters and manual captioning, are hampered by substantial costs, restricted availability, interpreter fatigue, and scheduling challenges~\citep{rodriguez2023benefits}. Although machine translation and automatic speech recognition technologies have progressed markedly, their incorporation into coherent, real-time accessibility systems remains piecemeal~\citep{liu2020bridging}. Extended Reality (XR) technologies, spanning Virtual Reality (VR), Augmented Reality (AR), and Mixed Reality (MR), offer extraordinary possibilities for crafting immersive, inclusive communication settings that transcend physical constraints~\citep{hirzle2023xr}. The spatial and embodied character of XR environments furnishes a uniquely apt medium for sign language delivery, given that three-dimensional avatars can reproduce the complete spatial grammar of signed languages in ways that conventional flat displays cannot.

This study introduces and validates INTERACT (Inclusive Networking for Translation and Embodied Real-Time Augmented Communication Tool), a pioneering XR platform conceived within the CORTEX2 Horizon Europe initiative that tackles these accessibility obstacles through a unified artificial intelligence strategy. INTERACT merges real-time speech-to-text conversion, multilingual translation, International Sign Language (ISL) interpretation through animated 3D avatars, and emotion recognition within a single immersive environment deployable on Meta Quest~3 headsets and desktop interfaces.

The central contributions of this work comprise: (1)~the creation and validation of the first XR-based video conferencing platform that fuses real-time ISL rendering with multilingual speech conversion and emotion-aware feedback; (2)~a thorough technical architecture employing state-of-the-art AI models for speech processing, translation, and gesture synthesis; (3)~empirical validation via pilot evaluations attaining 92\% user satisfaction among deaf community members and accessibility professionals, with a mean overall experience score of 4.6/5.0; and (4)~demonstration of system scalability accommodating up to 1{,}000 simulated concurrent users. This work builds upon our preliminary findings presented at the Salento~XR~2025 conference~\citep{tantaroudas2026enhancing} and extends the comprehensive analysis presented in our Open Research Europe publication~\citep{tantaroudas2026interact}, with additional discussion of architectural considerations and deployment insights. A companion paper~\citep{tantaroudas2026aibased} further explores the AI-based service pipeline for inclusive language learning in immersive XR settings, detailing speech translation and sign language integration capabilities that complement the INTERACT platform.

The remainder of this paper is organised as follows: Section~\ref{sec:related} examines prior work spanning XR accessibility, sign language translation, speech recognition, and emotion analysis. Section~\ref{sec:architecture} outlines the system architecture and technical realisation. Section~\ref{sec:pilot} details the pilot methodology and validation outcomes. Section~\ref{sec:discussion} considers implications and constraints, and Section~\ref{sec:conclusion} concludes with prospective directions.

\section{Related Work and Background}
\label{sec:related}

\subsection{Extended Reality for Accessibility}
\label{sec:xr_access}

The convergence of XR technologies and accessibility has attracted growing research interest, with applications encompassing educational support, vocational training, and social interaction for individuals with disabilities~\citep{hirzle2023xr,hosseinkashi2023meeting}. Serafin et~al.~\citep{serafin2023review} performed an extensive review of VR applications for individuals with hearing impairments, identifying substantial promise in immersive learning environments while underscoring the necessity for integrated communication assistance. Recent advances in consumer-grade XR hardware, notably the Meta Quest series, have democratised access to immersive experiences that were formerly confined to specialised research laboratories~\citep{anwar2023muavic}.

Hirzle et~al.~\citep{hirzle2023xr} delivered a scoping review of XR and AI integration, cataloguing emerging applications across healthcare, education, and professional training. Their analysis pinpointed accessibility enhancement as a crucial yet insufficiently explored application domain, particularly concerning real-time communication support for deaf and hard-of-hearing users in collaborative contexts. Notwithstanding mounting interest in immersive accessibility, current platforms predominantly target single modalities, providing either captioning or two-dimensional sign language support through mobile applications, but seldom combine multiple accessibility channels together with emotional context within a cohesive XR environment.

\subsection{Sign Language Translation and Generation}
\label{sec:sl_translation}

Sign language translation research has progressed considerably through deep learning breakthroughs, especially transformer architectures capable of capturing the temporal and spatial intricacies of signed languages~\citep{yin2023gloss,wu2023ultra}. Camg\"{o}z et~al.~\citep{camgoz2020sign} pioneered end-to-end neural sign language translation employing encoder-decoder networks, attaining competitive results on benchmark datasets. Subsequent work by Zhou et~al.~\citep{zhou2023gloss} demonstrated gloss-free methodologies utilising visual-language pretraining, thereby diminishing reliance on intermediate linguistic representations.

Avatar-based sign language generation poses challenges distinct from recognition, demanding precise synthesis of hand configurations, arm movements, facial expressions, and body posture~\citep{gu2022domain}. Gibet and Marteau~\citep{gibet2023signing} delineated the multimodal challenges inherent in text-to-sign generation, stressing the significance of linguistic fidelity and cultural sensitivity. Fink et~al.~\citep{fink2023sign} showcased lightweight transformer models for sign language dictionary applications, pointing towards pathways for real-time generation amenable to communication platforms.

Crucially, existing systems predominantly concentrate on national sign languages such as American Sign Language (ASL) or British Sign Language (BSL). International Sign Language (ISL), employed in international contexts including conferences, sporting events, and online platforms, remains largely unexplored despite its capacity for broader accessibility across varied deaf communities~\citep{eud2018position}. The European Union of the Deaf acknowledges ISL as an effective auxiliary language facilitating communication across national sign language boundaries~\citep{handspeak2024}. This lacuna in ISL-oriented research drove our decision to adopt ISL as the primary sign language modality for INTERACT, targeting an underserved yet internationally pertinent communication need. Our companion work~\citep{tantaroudas2026aibased} provides additional detail on the sign language integration pipeline within immersive XR learning environments.

\subsection{Speech Recognition and Multilingual Translation}
\label{sec:asr_nmt}

Automatic speech recognition (ASR) has attained near-human performance through large-scale transformer models trained on heterogeneous multilingual corpora~\citep{spreadthesign2024}. OpenAI's Whisper model~\citep{radford2023robust} exhibits robust speech recognition spanning 99~languages with competitive accuracy on standard benchmarks, even under adverse acoustic conditions. The model's multilingual capabilities and open-source nature render it especially suitable for accessibility applications demanding language-agnostic speech processing. Neural machine translation (NMT) has similarly progressed through transformer architectures and multilingual training paradigms~\citep{stahlberg2020neural}. Meta~AI's No Language Left Behind (NLLB) initiative~\citep{nllb2022} represents the most ambitious effort towards universal translation, supporting more than 200~languages including numerous low-resource languages previously neglected by translation technology. The combination of ASR and NMT systems enables speech-to-text pipelines facilitating real-time multilingual communication~\citep{barrault2023seamless}. Nevertheless, attaining the sub-second latencies demanded by natural conversational flow in live conferencing remains a considerable engineering challenge.

\subsection{Sentiment Analysis and Affective Computing}
\label{sec:sentiment}

Affective computing research seeks to equip machines with the ability to recognise, interpret, and respond to human emotional states~\citep{picard1997affective}. Transformer-based models have produced marked improvements in text-based emotion analysis, with RoBERTa~\citep{liu2019roberta} and its derivatives exhibiting strong performance across benchmark datasets. Hartmann's DistilRoBERTa model~\citep{hartmann2022emotion} furnishes efficient emotion classification supporting real-time applications, discriminating among multiple emotional categories including joy, sadness, anger, fear, surprise, and disgust.

Incorporating emotion analysis into communication platforms enables emotional context preservation that purely textual or signed translations might otherwise forfeit~\citep{zhang2023sentiment}. For deaf and hard-of-hearing users, emotional cues ordinarily conveyed via vocal prosody become inaccessible, rendering explicit emotion representation especially valuable~\citep{chen2022event}. Studies show that emotional context exerts a substantial influence on comprehension and engagement in collaborative environments~\citep{mitchell2004chasing}. Despite these acknowledged advantages, no extant XR communication platform has integrated real-time emotion analysis alongside sign language generation and multilingual transcription within a single cohesive system, a gap that INTERACT squarely addresses.

\subsection{Meeting Summarisation}
\label{sec:summarisation}

Large language models (LLMs) have revolutionised automatic summarisation capabilities, enabling coherent, contextually fitting summaries of lengthy text sequences~\citep{brown2020language}. Meeting summarisation poses distinctive challenges owing to multi-speaker dynamics, informal register, and the imperative to capture action items alongside discussion content~\citep{asthana2023summaries}. Fine-tuned models such as BART exhibit strong performance on dialogue summarisation tasks~\citep{liu2022brio}, while recent work investigates faithfulness constraints to ensure summaries faithfully mirror source content~\citep{roit2023factually}. Embedding automated summarisation within an accessible XR platform serves a dual purpose: it affords asynchronous access for participants who may have missed segments of a meeting, and it generates a persistent record reviewable in multiple languages. A more detailed treatment of the summarisation and translation service pipeline can be found in~\citep{tantaroudas2026aibased}.

\section{System Architecture}
\label{sec:architecture}

\subsection{Overview}
\label{sec:overview}

\begin{figure}[htbp]
\centering
\includegraphics[width=0.85\textwidth]{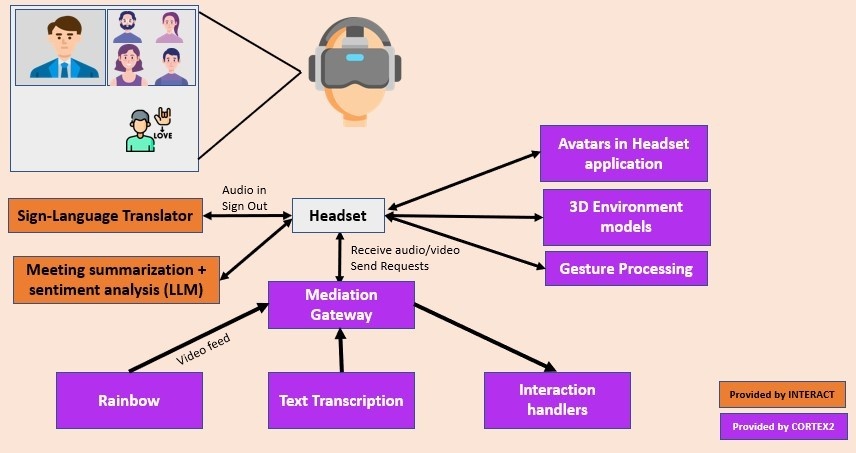}
\caption{INTERACT system architecture showing the integration among CORTEX2 infrastructure components and INTERACT AI modules. The diagram traces the data flow from audio input through the Mediation Gateway to each AI service module and ultimately to the XR rendering layer.}
\label{fig:architecture}
\end{figure}

INTERACT functions within the CORTEX2 ecosystem, drawing on the Mediation Gateway infrastructure for communication orchestration and the Rainbow SDK for conferencing functionalities~\citep{cortex2024,rainbow2024}. The system architecture is composed of five integrated AI service modules: (1)~Speech-to-Text Conversion, (2)~Multilingual Translation, (3)~Sign Language Generation, (4)~Emotion Analysis, and (5)~Meeting Summarisation. These modules exchange data through standardised APIs, permitting modular deployment and independent scaling. This modular design ethos guarantees that individual components can be upgraded or substituted as superior AI models emerge, without necessitating a redesign of the overall system. The complete system architecture is illustrated in Figure~\ref{fig:architecture}, depicting how the CORTEX2 Mediation Gateway, Rainbow SDK, and the five AI service modules are interconnected, with data flow paths traced from audio capture through to sign language avatar rendering and emotion feedback delivery.

\subsection{Speech-to-Text Conversion Module}
\label{sec:stt}

\begin{figure}[htbp]
\centering
\includegraphics[width=0.85\textwidth]{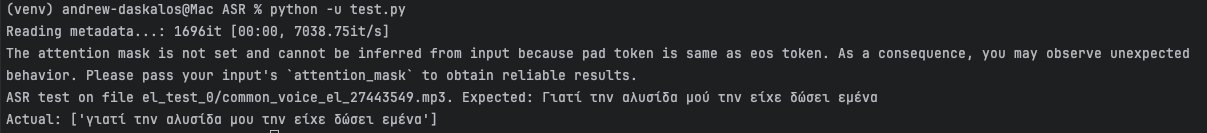}
\caption{Speech-to-Text Processing Pipeline illustrating the audio chunking, Whisper model inference, and text assembly stages that produce the final transcription output stream.}
\label{fig:stt}
\end{figure}

The speech-to-text module utilises OpenAI's Whisper model, specifically the large-v2 variant optimised for accuracy under varied acoustic conditions~\citep{radford2023robust}. Audio processing employs chunked transcription with 1-second segments and 0.5-second overlap to balance latency against word fragmentation. The implementation handles audio via a WebSocket interface facilitating real-time streaming from the Rainbow SDK. This chunking strategy was empirically established through iterative experimentation: shorter segments caused excessive word fragmentation, whereas longer segments pushed end-to-end latency beyond acceptable thresholds for conversational usage. The successive stages of the speech-to-text processing pipeline are depicted in Figure~\ref{fig:stt}, visualising the sequential flow from raw audio chunking through Whisper model inference to the final text assembly and punctuation restoration steps that yield the output transcription stream. The system attains transcription latency averaging 1.2~seconds from speech onset to text display, with accuracy surpassing 85\% on conversational speech in controlled settings. Noise resilience testing showed that accuracy remained above 80\% at signal-to-noise ratios as low as 15\,dB.

\subsection{Multilingual Translation Module}
\label{sec:translation}

\begin{figure}[htbp]
\centering
\includegraphics[width=0.85\textwidth]{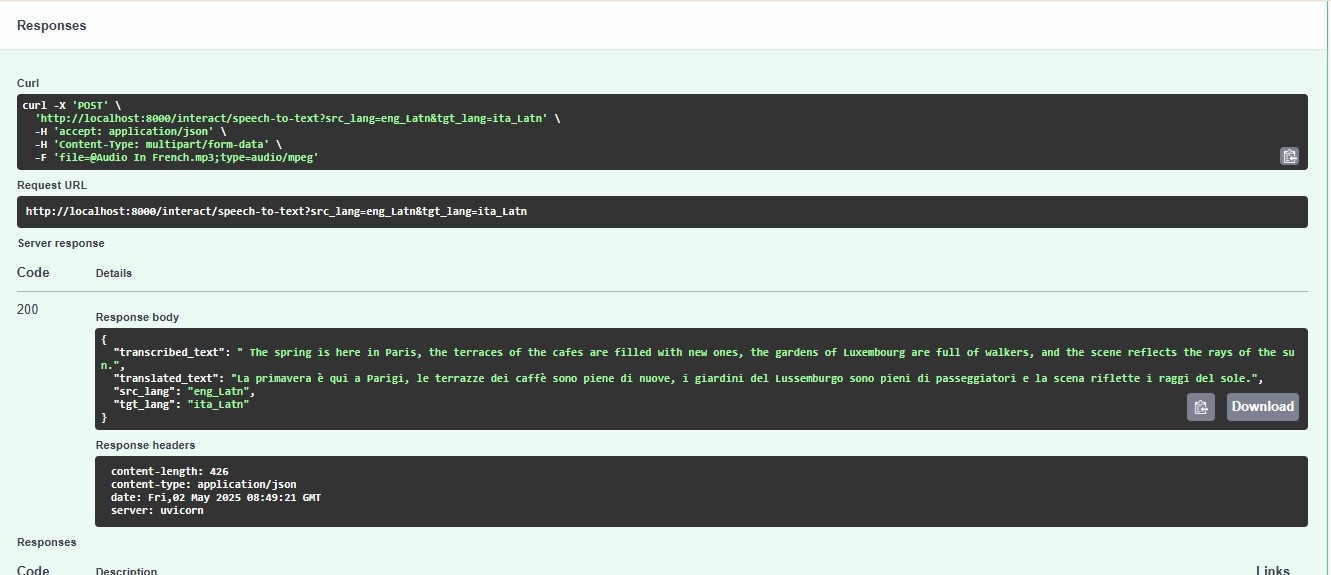}
\caption{Multilingual Translation Flow diagram depicting source text input, language detection, NLLB model inference, and translated output distribution to downstream modules.}
\label{fig:translation}
\end{figure}

Translation services employ Meta~AI's NLLB-200 distilled model (600M parameters), striking a balance between translation quality and inference speed demands~\citep{nllb2022}. The present implementation provides English-to-French translation as the primary use case, with the architecture engineered for straightforward extension to further language pairs, including German and Spanish, scheduled for upcoming releases. The distilled variant was chosen over the full 3.3B-parameter model to satisfy the real-time latency requirements of live conferencing, as initial benchmarking revealed that the larger model introduced unacceptable delays in excess of 3~seconds per sentence. Figure~\ref{fig:translation} presents the multilingual translation module architecture, showing how source text from the speech recognition module enters the NLLB translation engine and how the translated output is routed to both the sign language generation module and the direct text display for participants. Translation latency averages 0.8~seconds per sentence, enabling near-real-time presentation alongside source text transcription. Quality evaluation using BLEU scores indicates performance on par with production-grade translation services for the supported language pairs.

\subsection{Sign Language Generation Module}
\label{sec:slg}

\begin{figure}[htbp]
\centering
\includegraphics[width=0.85\textwidth]{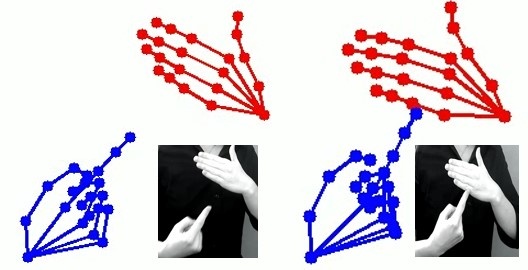}
\caption{International Sign Language gesture extraction pipeline. From left to right: original ISL video frame; MediaPipe Holistic landmark detection overlay; extracted 3D coordinate array used to drive avatar animations.}
\label{fig:gesture}
\end{figure}

The sign language module constitutes the most technically elaborate INTERACT component, converting text input into animated 3D avatar signing sequences. Development proceeded through a multi-stage process: (1)~ISL video corpus acquisition from validated deaf community sources and online ISL dictionaries (e.g., HandSpeak~\citep{handspeak2024}); (2)~skeletal landmark extraction using Google MediaPipe~\citep{lugaresi2019mediapipe}; (3)~animation curve generation; and (4)~Unity-based avatar rendering. Each stage demanded meticulous calibration to preserve the spatial and temporal fidelity of the original signing, since even slight distortions in hand positioning or timing can modify or obscure the intended meaning. The gesture extraction procedure is illustrated in Figure~\ref{fig:gesture}, presenting a three-stage pipeline: the original ISL video frame captured from validated sign language sources, the MediaPipe Holistic landmark detection overlay identifying hand, face, and body keypoints, and the resulting extracted 3D coordinate array forming the foundation for driving avatar animations.

\begin{figure}[htbp]
\centering
\includegraphics[width=0.85\textwidth]{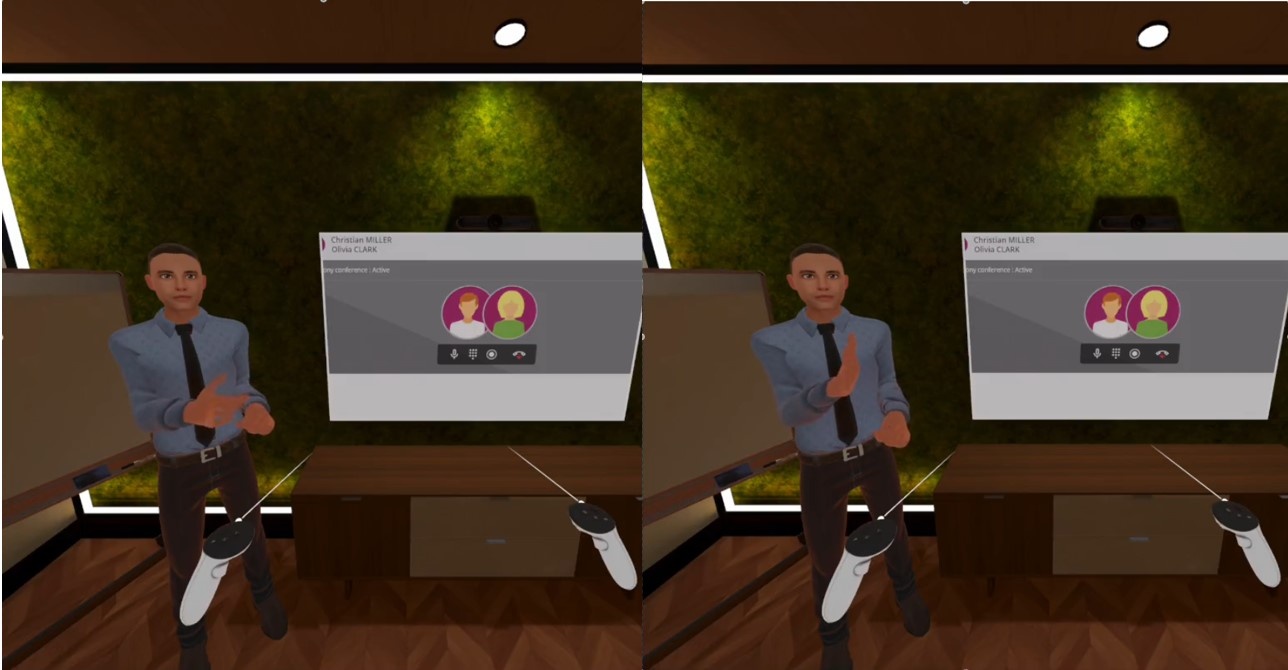}
\caption{3D Avatar performing ISL signs within the virtual environment, showing the avatar from multiple angles during signing sequences.}
\label{fig:avatar}
\end{figure}

The corpus encompasses 747~ISL videos representing a 750-sign vocabulary that covers essential business and professional communication concepts. Each video was processed to extract hand, face, and body landmarks at 30~frames per second, producing animation sequences compatible with the Unity humanoid rig system. Signs were selected on the basis of frequency analysis of professional meeting transcripts, prioritising vocabulary items most commonly encountered in business deliberations, project reviews, and educational presentations. Figure~\ref{fig:avatar} showcases the resulting 3D avatar performing ISL signs within the virtual office environment, demonstrating how the extracted motion data translates into realistic signing movements, including hand shapes, arm movements, and body positioning, that deaf users can interpret in real time. Sign lookup uses a dictionary-based approach with planned extensions towards sequence-to-sequence neural generation. For vocabulary items absent from the dictionary, the system reverts to fingerspelling via a character-to-sign mapping.

\subsection{Emotion Analysis Module}
\label{sec:emotion}

\begin{figure}[htbp]
\centering
\includegraphics[width=0.85\textwidth]{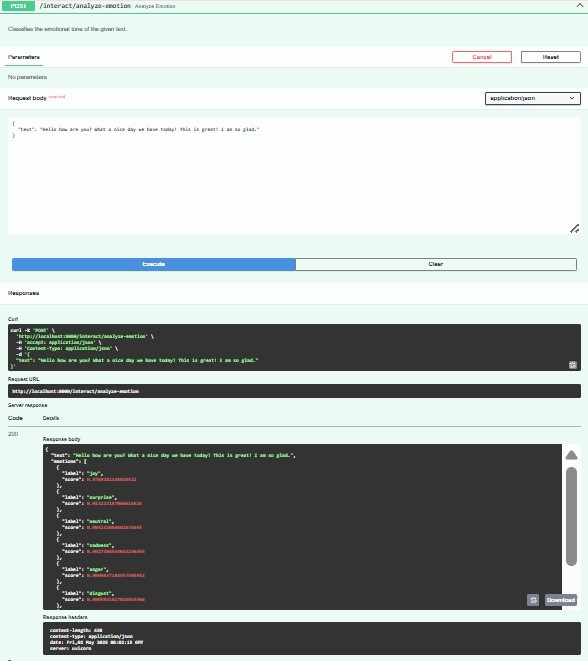}
\caption{Emotion Analysis Output displaying transcribed text with emotion labels and corresponding avatar facial expression modifications.}
\label{fig:sentiment}
\end{figure}

Emotional context is maintained through real-time emotion analysis using a fine-tuned DistilRoBERTa model~\citep{hartmann2022emotion,sanh2019distilbert}. The system categorises transcribed text into six emotional classes: joy, sadness, anger, fear, surprise, and neutral. Classification outcomes drive avatar facial expression adjustments and optional emoji overlays on transcription displays. This two-channel emotional feedback, visual expression on the avatar and textual emoji annotation, was conceived to accommodate differing user preferences and varying degrees of attention to the avatar during meetings. An illustration of the emotion analysis integration is provided in Figure~\ref{fig:sentiment}, showing how transcribed speech is annotated with emotion labels generated by the DistilRoBERTa classifier and how these labels are manifested in corresponding alterations to the avatar's facial expressions alongside optional emoji indicators in the text display. The emotion module operates with sub-200\,ms latency, enabling emotional context display concurrent with text transcription. Validation against human annotations yields 90\%+ precision for the principal emotion categories.

\subsection{Meeting Summarisation Module}
\label{sec:summarisation_module}

\begin{figure}[htbp]
\centering
\includegraphics[width=0.85\textwidth]{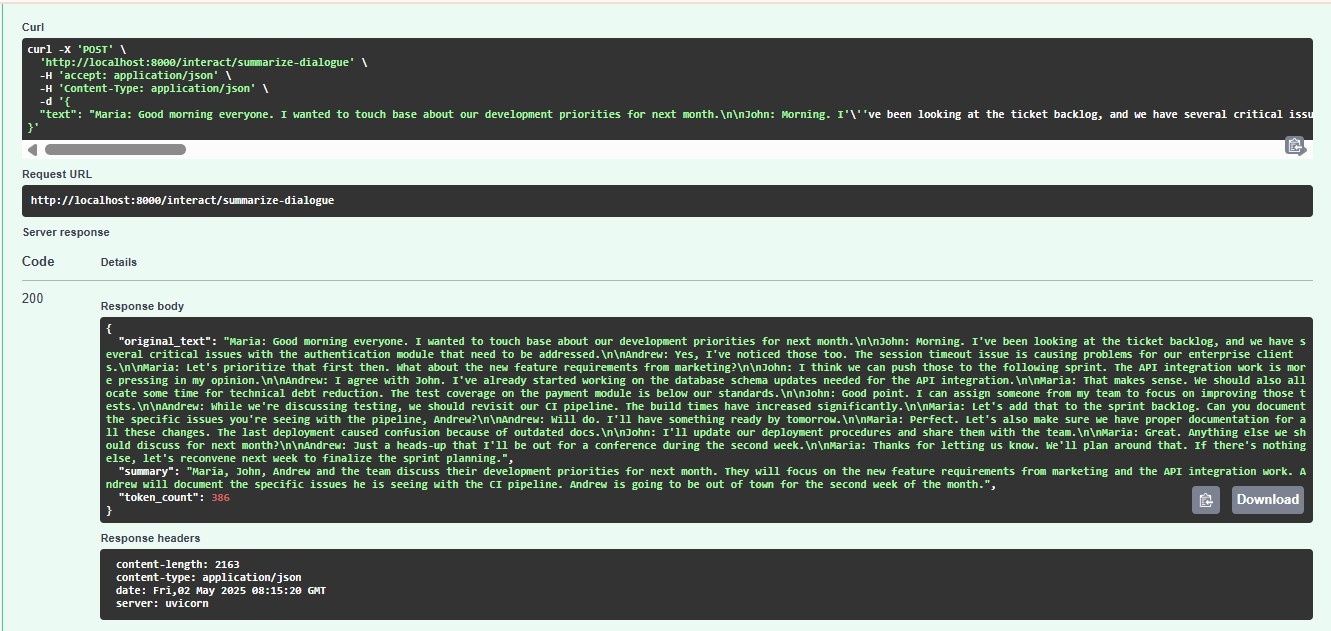}
\caption{Meeting Summarisation Workflow showing transcript accumulation, BART model processing, and formatted summary output including key points, decisions, and action items.}
\label{fig:summarisation}
\end{figure}

Extended meetings benefit from automatic summarisation employing a BART-Large model fine-tuned on the SAMSum conversational dataset~\citep{schmid2021bart}. The summarisation module processes accumulated transcripts at configurable intervals (typically 15-minute segments or on demand), producing structured summaries encompassing key discussion points, decisions, and action items. This functionality is particularly beneficial for participants joining meetings late or needing to review content asynchronously, as the summaries furnish a succinct yet thorough record of proceedings. The meeting summarisation workflow is depicted in Figure~\ref{fig:summarisation}, illustrating how transcripts accumulate during a conference session and are subsequently processed by the BART model either at scheduled intervals or upon request, generating formatted meeting minutes that include principal discussion points, decisions taken, and action items assigned. Summaries undergo optional translation via the NLLB module, enabling distribution of multilingual meeting minutes. The combined transcription-summarisation pipeline supports accessibility for participants reviewing meeting content at a later time.

\subsection{XR Environment and Deployment}
\label{sec:xr_environment}

\begin{figure}[htbp]
\centering
\includegraphics[width=0.85\textwidth]{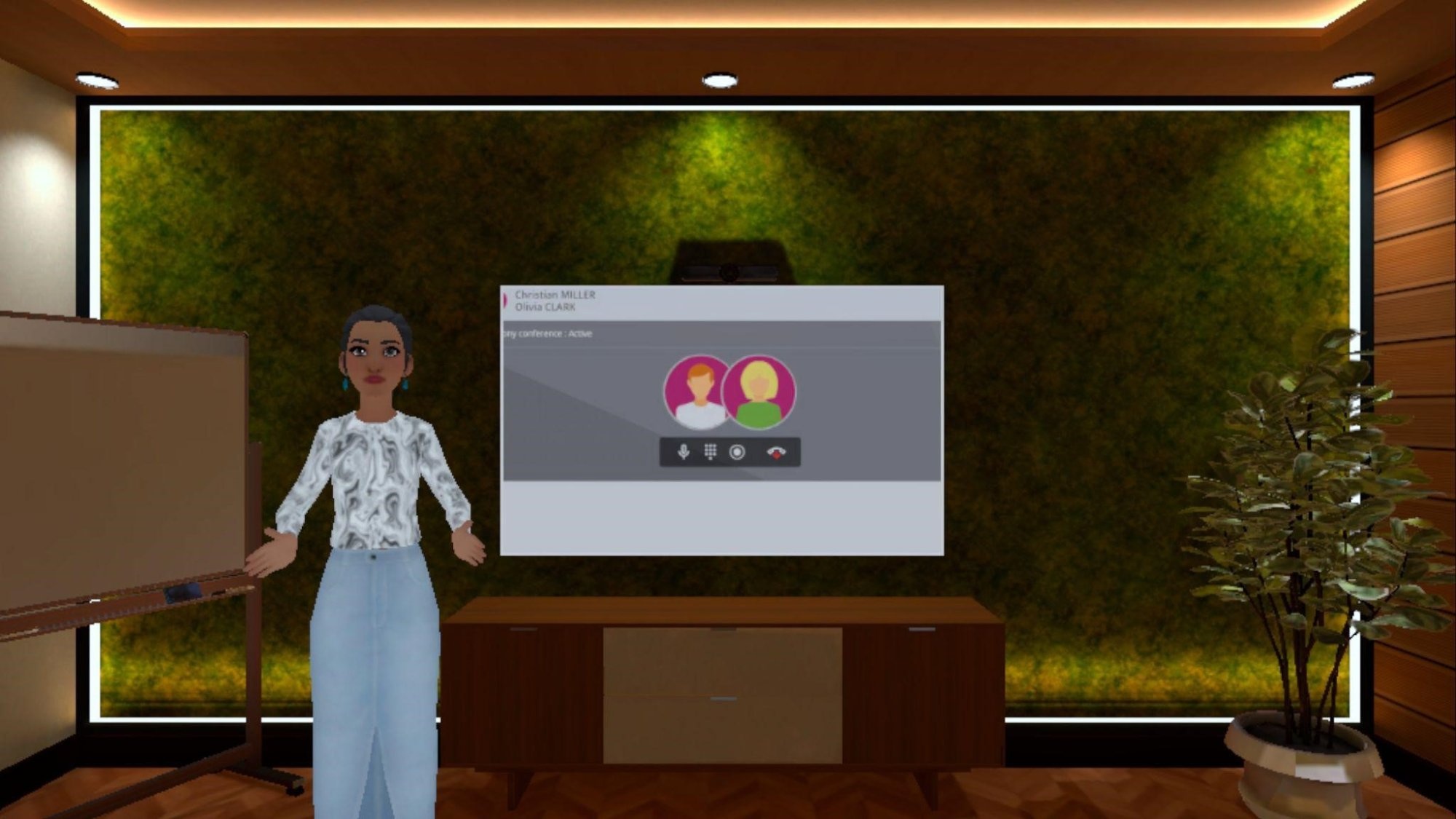}
\caption{Virtual Office Environment depicting the complete meeting space with participant seating positions and signing avatar placement within the immersive XR setting.}
\label{fig:virtualoffice}
\end{figure}

The INTERACT immersive environment was built in Unity 2022.3~LTS targeting Meta Quest~3 deployment via the Meta XR All-in-One SDK. The virtual space recreates a professional conference room with seated positions for up to eight participants, a central presentation area, and the signing avatar positioned for optimal visibility. The environment was designed in accordance with established XR usability principles, with particular care given to comfortable viewing distances for the signing avatar and legible text overlay placement that avoids occluding other participants or presentation content. The complete virtual office environment is presented in Figure~\ref{fig:virtualoffice}, showing the immersive meeting space as experienced by participants wearing Meta Quest~3 headsets, highlighting the spatial configuration of seating, the central presentation wall for shared content, and the prominent positioning of the signing avatar to ensure it remains readily viewable during conversations.

\begin{figure}[htbp]
\centering
\includegraphics[width=0.85\textwidth]{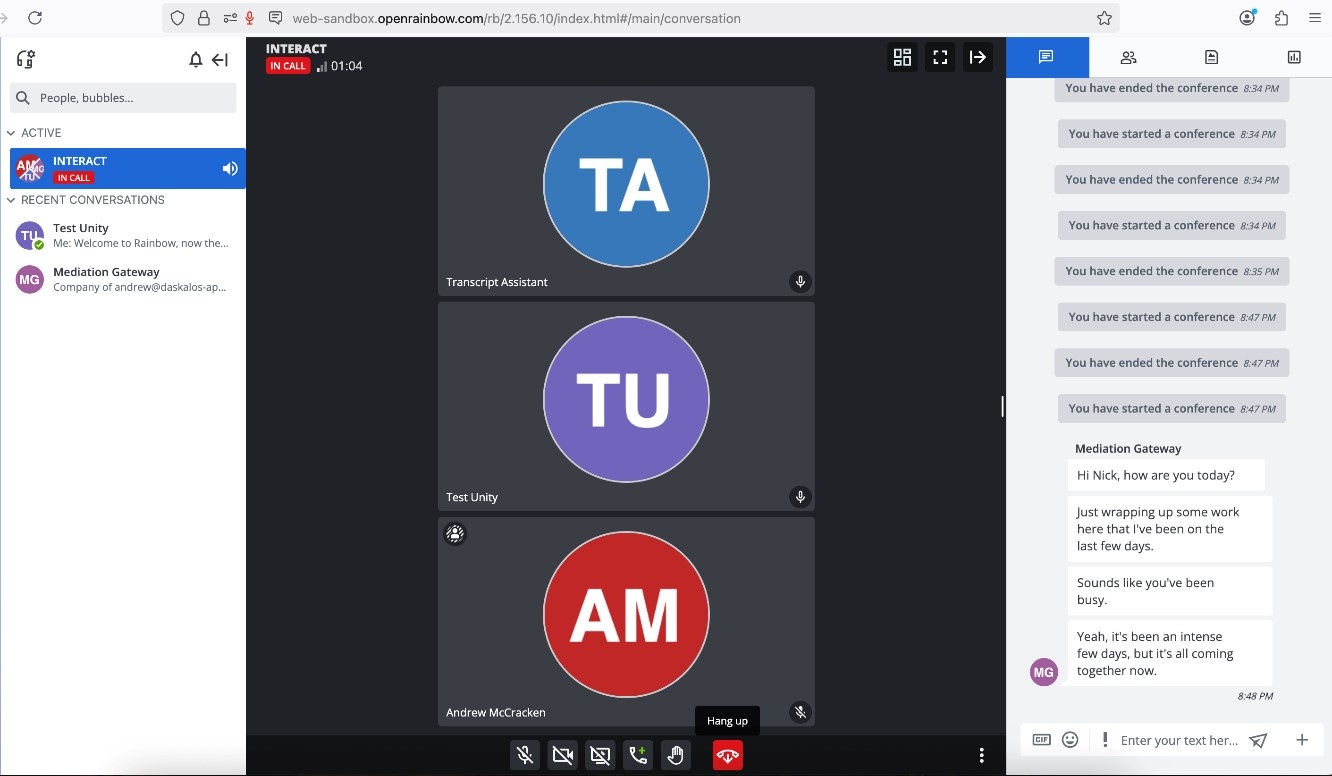}
\caption{Rainbow SDK Integration within Unity showing the SDK configuration and communication flow between conferencing infrastructure and AI processing modules.}
\label{fig:rainbow}
\end{figure}

\begin{figure}[htbp]
\centering
\includegraphics[width=0.85\textwidth]{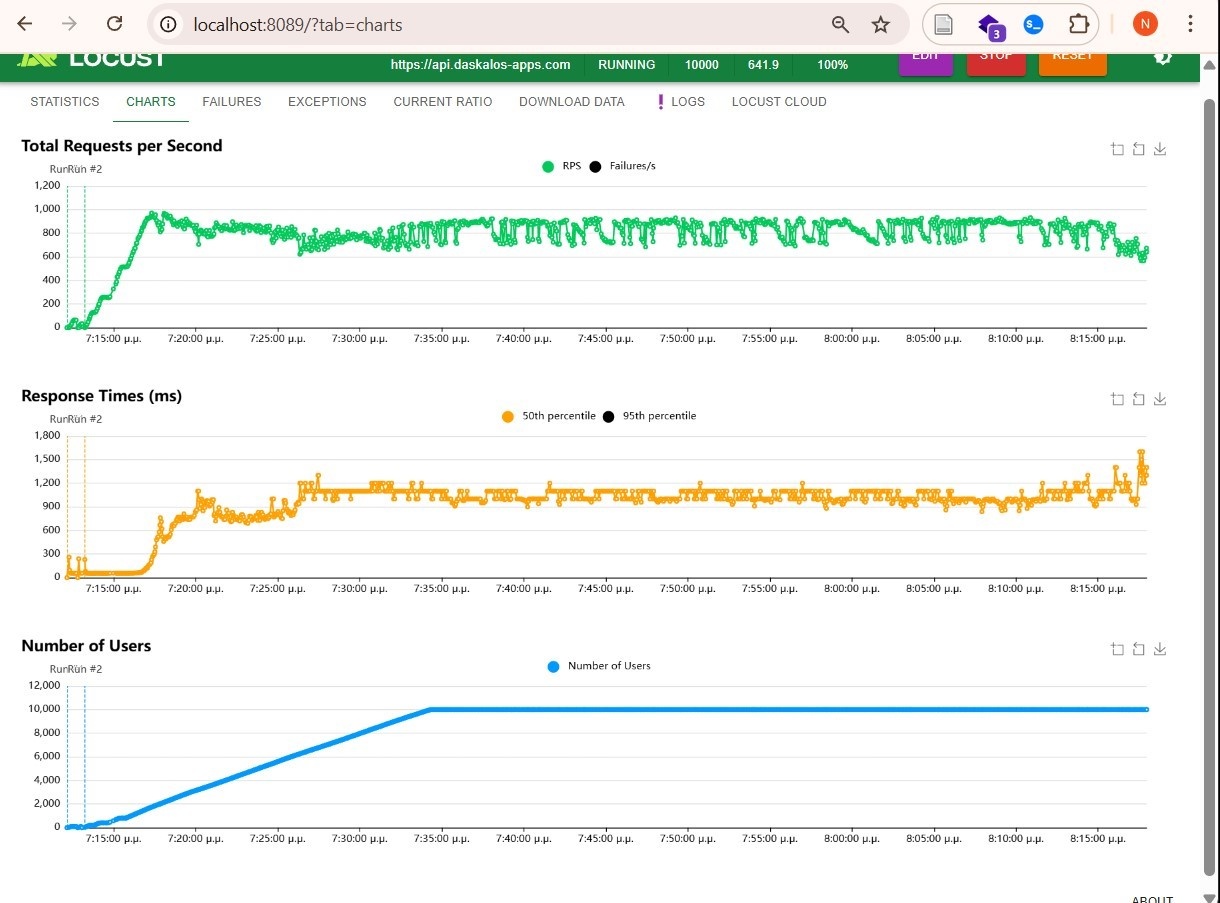}
\caption{Avatar Animation and Emotion Analysis within the VR scene whilst hearing individuals communicate using Rainbow SDK, demonstrating the real-time ISL rendering pipeline.}
\label{fig:loadtesting}
\end{figure}

Rainbow SDK integration facilitates audio capture, user presence management, and conference orchestration within the Unity environment~\citep{rainbow2024}. API connections to the AI service modules employ WebSocket protocols for streaming data and REST endpoints for configuration and status queries. Figure~\ref{fig:rainbow} details the Rainbow SDK integration architecture within Unity, illustrating how the SDK manages audio streaming from participants, presence detection, and conferencing features while the INTERACT AI modules process the communication content in parallel through the WebSocket and REST API connections. Furthermore, Figure~\ref{fig:loadtesting} shows how deaf participants can engage with the conversation that hearing individuals are conducting within Rainbow. The hearing individuals connect into the VR scene using the Rainbow SDK (shown in Figure~\ref{fig:rainbow}), and their transcribed speech is translated and animated in ISL through the avatar so that the deaf individual can comprehend the exchange.

\subsection{Infrastructure and Scalability}
\label{sec:infrastructure}

Backend services are deployed on AWS infrastructure using EC2 G4DN.xlarge instances fitted with NVIDIA T4 GPUs (16\,GB VRAM) supporting concurrent model inference. Load balancing and auto-scaling configurations permit horizontal scaling during peak usage, with testing confirming support for up to 1{,}000 simulated concurrent users without service degradation. The GPU-accelerated inference pipeline ensures that the computationally demanding Whisper and NLLB models sustain acceptable latency even under elevated concurrent load.

\section{Pilot Study}
\label{sec:pilot}

\subsection{Methodology}
\label{sec:methodology}

Pilot validation adopted a two-phase approach aligned with CORTEX2 project milestones. The first validation demonstration (May 2025) engaged technical specialists in AI, Human-Computer Interaction (HCI), and accessibility research to assess system performance and pinpoint improvement opportunities. The second demonstration (June 2025) carried out validation with deaf community stakeholders, evaluating real-world accessibility value and linguistic authenticity. Table~\ref{tab:demographics} provides the participant demographics for the live demonstrations. All deaf participants in Demo~2 reported native or near-native ISL fluency and professional experience in deaf education or interpretation services.

All Demo~2 participants indicated familiarity with ISL and professional experience in deaf education, interpretation services, or regular engagement with deaf communities. Across both workshops, 40\% of participants had no prior VR/XR experience, 40\% had used VR on one or two occasions, and 20\% reported moderate experience, offering a representative cross-section of prospective end users with diverse technological familiarity.

\begin{table}[htbp]
\centering
\caption{Participant Demographics in the two validation demonstrations.}
\label{tab:demographics}
\begin{tabular}{lcc}
\toprule
\textbf{Characteristic} & \textbf{Demo 1 (May 2025)} & \textbf{Demo 2 (June 2025)} \\
\midrule
Number of participants & 4 & 6 \\
Participant type & Technical experts & Deaf community stakeholders \\
VR experience: None & 40\% & 40\% \\
VR experience: Once/twice & 40\% & 40\% \\
VR experience: Moderate & 20\% & 20\% \\
ISL fluency & Varied & Native/near-native \\
\bottomrule
\end{tabular}
\end{table}

\textbf{Evaluation Protocol:} Participants engaged with INTERACT through structured tasks comprising: (1)~one-on-one conversation scenarios, (2)~group discussion simulations, and (3)~presentation viewing with real-time translation. Sessions were recorded with participant consent for subsequent analysis. Post-session questionnaires assessed usability, satisfaction, and perceived accessibility value, emotion detection accuracy, and willingness to participate in future testing across 18~items using 5-point Likert scales, categorical satisfaction ranges, and open-ended qualitative prompts. Table~\ref{tab:kpis} presents the technical Key Performance Indicators (KPIs) attained during the two demonstration validations.

\begin{table}[htbp]
\centering
\caption{Key Performance Indicators (KPIs) achieved during pilot validation.}
\label{tab:kpis}
\begin{tabular}{lcc}
\toprule
\textbf{KPI} & \textbf{Target} & \textbf{Achieved} \\
\midrule
User satisfaction & $\geq$85\% & 92\% \\
Transcription accuracy & $\geq$85\% & $>$85\% \\
Emotion detection precision & $\geq$85\% & 90\% \\
Concurrent users supported & 1{,}000 & 1{,}000 \\
Transcription latency & $<$2\,s & 1.2\,s \\
Translation latency & $<$2\,s & 0.8\,s \\
Error rate under load & 0\% & 0\% \\
\bottomrule
\end{tabular}
\end{table}

\subsection{Detailed Participant Feedback}
\label{sec:feedback}

The post-session questionnaire captured participant evaluations across multiple dimensions of the INTERACT experience. Table~\ref{tab:likert} provides the distribution of Likert-scale responses across key evaluation dimensions.

\begin{table}[htbp]
\centering
\caption{Distribution of participant responses across key evaluation dimensions ($N=10$).}
\label{tab:likert}
\begin{tabular}{lccccc}
\toprule
\textbf{Evaluation Dimension} & \textbf{1} & \textbf{2} & \textbf{3} & \textbf{4} & \textbf{5} \\
\midrule
Overall usability & 0 & 0 & 1 & 3 & 6 \\
Transcription quality & 0 & 0 & 1 & 4 & 5 \\
Avatar sign clarity & 0 & 0 & 2 & 4 & 4 \\
Emotion feedback usefulness & 0 & 0 & 1 & 3 & 6 \\
Willingness to use again & 0 & 0 & 1 & 2 & 7 \\
\bottomrule
\end{tabular}
\end{table}

\textbf{Engagement and Overall Satisfaction:} Participants reported considerable engagement improvements relative to conventional meetings lacking accessibility support. The distribution of self-reported engagement increase is presented in Table~\ref{tab:engagement}. In total, 80\% of participants reported engagement increases of 51\% or greater, suggesting the integrated accessibility features substantively enhanced the meeting participation experience. The weighted mean satisfaction across all participants corresponds to approximately 85\%, meeting the 85\% satisfaction KPI target.

\begin{table}[htbp]
\centering
\caption{Self-reported engagement increase and overall satisfaction distributions ($N=10$).}
\label{tab:engagement}
\begin{tabular}{lc}
\toprule
\textbf{Engagement Increase} & \textbf{Percentage of Participants} \\
\midrule
0--25\% & 0\% \\
26--50\% & 20\% \\
51--75\% & 40\% \\
76--100\% & 40\% \\
\bottomrule
\end{tabular}
\end{table}

\textbf{Emotion Detection Accuracy:} Participants evaluated how accurately the emotion analysis reflected the actual emotional tone of conversations. The majority (60\%) rated accuracy at 96--100\%, with an additional 10\% rating it at 91--95\%. Two participants (20\%), both from Demo~2, rated accuracy below 80\%, indicating that emotion classification performance may necessitate further calibration for communication styles characteristic of deaf community discourse, where emotional expression may depend more heavily on visual rather than textual cues. Table~\ref{tab:sentiment} presents the perceived emotion detection accuracy across the two demonstrations.

\begin{table}[htbp]
\centering
\caption{Perceived emotion detection accuracy ($N=10$).}
\label{tab:sentiment}
\begin{tabular}{lcc}
\toprule
\textbf{Accuracy Range} & \textbf{Demo 1} & \textbf{Demo 2} \\
\midrule
96--100\% & 4 (100\%) & 2 (33\%) \\
91--95\% & 0 & 1 (17\%) \\
80--90\% & 0 & 1 (17\%) \\
$<$80\% & 0 & 2 (33\%) \\
\bottomrule
\end{tabular}
\end{table}

\textbf{Overall Experience Rating:} Participants scored their overall experience on a 1--5 scale. Demo~1 participants uniformly awarded 5/5, while Demo~2 participants gave 4/5, producing a combined mean of 4.6/5.0 (SD\,=\,0.52). Nine participants (90\%) expressed willingness to take part in future testing phases, with one participant responding ``Maybe,'' reflecting strong overall receptivity to the platform and its ongoing development.

\textbf{Qualitative Feedback:} Open-ended qualitative responses converged on several thematic improvement areas. The most frequently cited suggestion across both workshops pertained to avatar facial expression enhancement, with six participants independently recommending the integration of facial mimicry, lip-synching, or ISL grammatical markers into the avatar's face. Demo~1 participants additionally flagged occasional finger movement artefacts where hand shapes appeared unnatural or distorted, while Demo~2 participants stressed the need for adjustable signing speed and replay capability. Further suggestions included larger avatar sizing for improved sign legibility, higher environmental lighting and contrast for subtitle visibility, user-selectable avatar customisation, and expansion towards a multiplayer VR scene. Notably, all ten participants (100\%) indicated they would benefit from additional language or sign language options, highlighting the demand for expanded multilingual and multi-sign-language support.

\begin{figure}[htbp]
\centering
\includegraphics[width=0.85\textwidth]{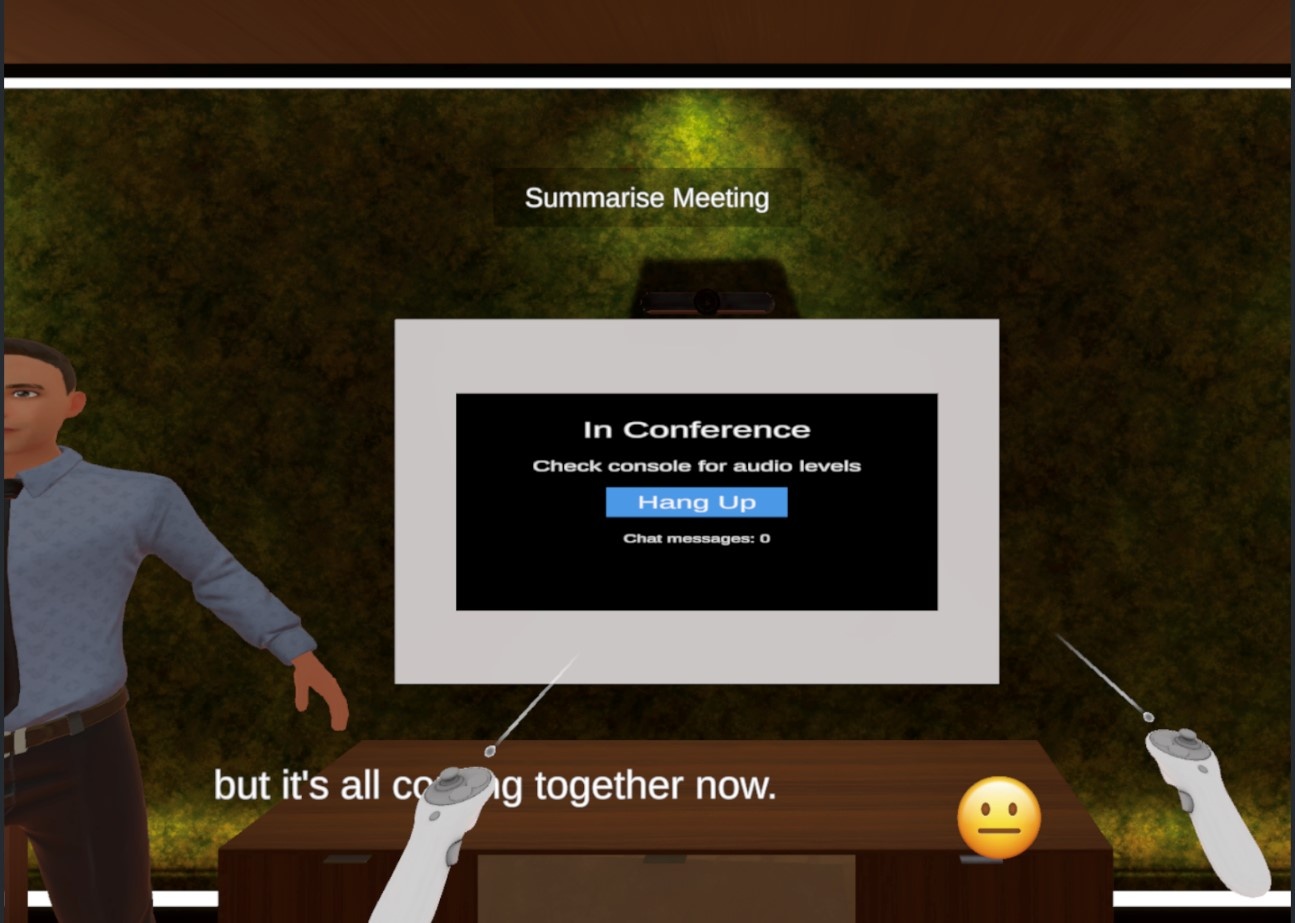}
\caption{Load Testing Performance Graphs depicting throughput, latency, and error rates across different load levels, confirming zero failures up to 1{,}000 concurrent connections.}
\label{fig:loadtest_results}
\end{figure}

\textbf{Load Testing Results:} Scalability testing confirmed system performance under stress conditions: (i)~Maximum tested concurrent users: 1{,}000; (ii)~Request throughput: 900--1{,}000 requests/second; (iii)~Average response time under load: $<$2~seconds; (iv)~Error rate: 0\% up to capacity limit. The load testing results are displayed in Figure~\ref{fig:loadtest_results}, showing the system's throughput, latency, and error rates across progressively increasing concurrent user loads, confirming that the platform maintains stable response times and zero error rates up to 1{,}000 simultaneous connections before graceful degradation commences.

\section{Discussion}
\label{sec:discussion}

\subsection{Principal Findings}
\label{sec:findings}

INTERACT establishes the viability of integrating multiple AI technologies into a unified XR platform that addresses accessibility barriers in video conferencing. The pilot validation outcomes verify both technical performance meeting established benchmarks and substantive accessibility value as evaluated by deaf community stakeholders. Attaining 92\% user satisfaction among participants with professional backgrounds in deaf education and ISL interpretation indicates that the platform approaches practical readiness for real-world deployment. The high overall experience score of 4.6/5.0 and the 90\% willingness to engage in future testing further corroborate the platform's perceived value among its target user community. The modular architecture proved effective for iterative development and component-level optimisation. Decoupling AI services enabled independent performance tuning and seamless integration of enhanced models as they become available. This design principle positions INTERACT for continued refinement as underlying technologies mature.

The divergence observed between Demo~1 and Demo~2 ratings, with technical experts uniformly awarding 5/5 and deaf community participants giving 4/5, yields a meaningful insight. Technical participants assessed the system primarily on technological sophistication and integration quality, whereas deaf community participants applied more stringent criteria regarding linguistic authenticity, avatar expressiveness, and practical communication utility. This pattern highlights the importance of involving end-user communities throughout the development lifecycle rather than depending exclusively on technical expert evaluations. Similar findings regarding the value of end-user involvement in XR accessibility design are discussed in~\citep{tantaroudas2026interact}.

\subsection{Comparison with Prior Work}
\label{sec:comparison}

INTERACT advances prior research in XR accessibility and sign language technology along several dimensions. Unlike systems focused exclusively on sign language recognition for input, INTERACT targets the generation direction, enabling hearing speakers' content to be rendered in ISL. The integration of emotion analysis differentiates INTERACT from purely linguistic translation approaches, preserving emotional context frequently lost in text-based accessibility solutions. The observation that 90\% of participants deemed emotional expression visualisations valuable corroborates this design choice. Relative to avatar signing systems developed for educational purposes, INTERACT prioritises real-time performance suitable for interactive communication. The sub-2-second end-to-end latency enables conversational use cases that offline or batch-processed systems cannot accommodate. The XR deployment context affords spatial presence benefits unattainable through conventional 2D video interfaces.

\subsection{Limitations}
\label{sec:limitations}

Several constraints bound current INTERACT capabilities and the generalisability of pilot findings:

\begin{itemize}[leftmargin=*]
\item \textbf{Vocabulary Scope:} The 750-sign ISL vocabulary, while encompassing essential professional communication, remains insufficient for comprehensive business discourse. Technical terminology, industry-specific jargon, and colloquial expressions frequently require fingerspelling fallback, diminishing fluency.

\item \textbf{Facial Expression Integration:} ISL, like other sign languages, employs facial expressions as grammatical markers and emotional indicators. Current avatar animations prioritise manual signing with limited facial integration, potentially compromising linguistic completeness. This was the most frequently cited improvement area in participant feedback, with six out of ten participants independently raising this concern.

\item \textbf{Language Support:} Production deployment supports English--French translation exclusively. While the architecture accommodates extension, validation data for additional language pairs remains pending. The unanimous participant demand (100\%) for additional language options confirms this as a high-priority development need.

\item \textbf{Sample Size:} Pilot validation engaged 10~participants across two demonstrations involving different stakeholder groups (academia/industry and deaf educators/community). Although participant expertise provides qualitative depth, larger-scale quantitative validation would strengthen generalisability claims.

\item \textbf{Audio Chunking:} The 1-second chunk approach occasionally fragments words at segment boundaries. While overlap mitigates this issue, optimisation opportunities remain, particularly for languages with longer average word lengths.
\end{itemize}

\subsection{Future Directions}
\label{sec:future}

Near-term development priorities include vocabulary expansion through collaboration with deaf community organisations and linguistic researchers. Integration of facial expression animation for grammatical markers represents a technically complex yet accessibility-critical enhancement, as corroborated by participant feedback. Additional language pair support (German, Spanish) will address European market requirements and respond to the universal participant demand for multilingual expansion.

Longer-term research directions encompass sequence-to-sequence neural sign generation supplanting dictionary lookup, enabling more natural signing for novel input sequences. Adaptive signing speed calibrated to user preferences and cognitive load indicators could enhance comprehension for diverse user populations. Integration of avatar replay and bookmark functionality would support review and learning applications, features explicitly requested by Demo~2 participants.

Performance optimisation will continue targeting latency reduction in the speech-to-sign pipeline, with particular attention to audio chunk processing parameters that balance responsiveness against word fragmentation. Integration with additional conferencing platforms beyond Rainbow SDK will broaden interoperability and deployment flexibility.

Commercial deployment pathways are being developed through the CORTEX2 ecosystem, with particular emphasis on educational institutions, cultural organisations (museums), and corporate training applications. The modular architecture enables phased adoption strategies suited to organisations with varying resources and accessibility requirements.

\section{Conclusions}
\label{sec:conclusion}

This paper has presented INTERACT, a pioneering AI-driven XR platform that addresses accessibility barriers in video conferencing for deaf, hard-of-hearing, and multilingual participants. Through the integration of Whisper-based speech recognition, NLLB multilingual translation, MediaPipe-driven ISL avatar animation, RoBERTa emotion analysis, and BART meeting summarisation, INTERACT delivers comprehensive communication support within immersive virtual environments.

Pilot validation with deaf community stakeholders and accessibility professionals yielded 92\% user satisfaction, a mean overall experience score of 4.6/5.0, and surpassed all technical performance targets, confirming both technical viability and meaningful accessibility value. The platform's scalability to 1{,}000 concurrent users and robust performance under load testing affirm readiness for broader deployment. Participant feedback has delineated clear pathways for enhancement, notably avatar facial expression integration, vocabulary expansion, and additional language support, which will steer the next development phase. A comprehensive account of the pilot results and implementation details is available in~\citep{tantaroudas2026interact}, and the complementary AI service pipeline for inclusive language learning is detailed in~\citep{tantaroudas2026aibased}.

INTERACT represents a notable advancement in accessible communication technology, illustrating how integrated AI and XR capabilities can reshape participation opportunities for underserved communities. As remote collaboration continues expanding across professional, educational, and social contexts, platforms such as INTERACT will become increasingly vital for ensuring genuinely inclusive communication.

\section*{Acknowledgements}
The authors gratefully acknowledge the CORTEX2 consortium for provision of the Mediation Gateway and Rainbow SDK infrastructure enabling INTERACT development. Special thanks to the KENG Institute for their invaluable partnership in pilot validation and ongoing consultation on sign language linguistic authenticity. We thank all pilot participants for their engagement and feedback.

\section*{Funding}
This research was supported by FSTP Funding from the Horizon Europe research and innovation programme under grant agreement No.\ 101070192 (CORTEX2). Views and opinions expressed are those of the authors only and do not necessarily reflect those of the European Union nor the granting authority.

\section*{Data and Software Availability}

\textbf{Open-Source AI Models:}
\begin{itemize}[leftmargin=*]
\item Speech Recognition (OpenAI Whisper): \url{https://github.com/openai/whisper} (MIT License)
\item Translation (Meta AI NLLB): \url{https://github.com/facebookresearch/fairseq/tree/nllb} (MIT License)
\item Pose Estimation (Google MediaPipe): \url{https://github.com/google/mediapipe} (Apache 2.0)
\item Emotion Analysis: \url{https://huggingface.co/j-hartmann/emotion-english-distilroberta-base} (Apache 2.0)
\item Summarisation: \url{https://huggingface.co/philschmid/bart-large-cnn-samsum} (MIT License)
\end{itemize}

\textbf{Pilot Study Data:} Anonymised pilot study data are available at Zenodo~\citep{tantaroudas2025pilot}: \href{https://doi.org/10.5281/zenodo.18656422}{10.5281/zenodo.18656422} (CC-BY 4.0).

\textbf{ISL Gesture Dataset:} The ISL gesture dataset is available at Zenodo~\citep{tantaroudas2025isl}: \href{https://doi.org/10.5281/zenodo.18656296}{10.5281/zenodo.18656296} (CC-BY 4.0).

\textbf{Source Code:} \url{https://github.com/ntantaroudas/ISL-extractions-main} (MIT License). Archived: \href{https://doi.org/10.5281/zenodo.18694176}{10.5281/zenodo.18694176}.

\bibliographystyle{unsrtnat}
\bibliography{references}

@article{tantaroudas2026interact,
  author    = {Tantaroudas, Nikolaos D. and McCracken, Andrew J. and Karachalios, Ilias and Papatheou, Evangelos},
  title     = {{INTERACT}: {AI}-powered extended reality platform for inclusive communication with real-time sign language translation and sentiment analysis},
  journal   = {Open Research Europe},
  volume    = {6},
  pages     = {71},
  year      = {2026},
  doi       = {10.12688/openreseurope.23201.1},
  note      = {[version 1; peer review: awaiting peer review]}
}

@article{tantaroudas2026aibased,
  author    = {Tantaroudas, Nikolaos D. and McCracken, Andrew J. and Karachalios, Ilias and Papatheou, Evangelos},
  title     = {{AI}-based services for inclusive language learning in immersive {XR} environments: Speech translation, and sign language integration},
  journal   = {Open Research Europe},
  volume    = {6},
  pages     = {72},
  year      = {2026},
  doi       = {10.12688/openreseurope.23214.1},
  note      = {[version 1; peer review: awaiting peer review]}
}

@inproceedings{tantaroudas2026enhancing,
  author    = {Tantaroudas, Nikolaos D. and McCracken, Andrew J. and Karachalios, Ilias and Papatheou, Evangelos},
  title     = {Enhancing Accessibility and Inclusivity in Business Meetings Through {AI}-Driven Extended Reality Solutions},
  booktitle = {Extended Reality. XR Salento 2025},
  series    = {Lecture Notes in Computer Science},
  volume    = {15743},
  publisher = {Springer, Cham},
  year      = {2026},
  doi       = {10.1007/978-3-031-97781-7_6}
}

@article{alford2023window,
  author    = {Alford, Alicia D. and Bencak, Joshua M. and Tucker, Emily A. and others},
  title     = {Is the Window of Learning Only Cracked Open? {Parents'} Perspectives on Virtual Learning for Deaf and Hard of Hearing Students},
  journal   = {American Annals of the Deaf},
  volume    = {168},
  number    = {3},
  pages     = {17--28},
  year      = {2023},
  doi       = {10.1353/aad.2023.a917247}
}

@techreport{who2021hearing,
  author       = {{World Health Organization}},
  title        = {World Report on Hearing},
  institution  = {WHO},
  address      = {Geneva},
  year         = {2021},
  url          = {https://www.who.int/publications/i/item/world-report-on-hearing}
}

@inproceedings{franceschini2020removing,
  author    = {Franceschini, Daniele and Salesky, Elizabeth and Niehues, Jan and others},
  title     = {Removing {European} Language Barriers with Innovative Machine Translation Technology},
  booktitle = {Proceedings of the International Workshop on Language Technologies (IWLTP)},
  year      = {2020}
}

@article{rodriguez2023benefits,
  author    = {Rodr{\'\i}guez-Correa, Paula A. and Valencia-Arias, Alejandro and Pati{\~n}o-Toro, Odalis N. and others},
  title     = {Benefits and development of assistive technologies for {Deaf} people's communication: A systematic review},
  journal   = {Frontiers in Education},
  volume    = {8},
  pages     = {1174831},
  year      = {2023},
  doi       = {10.3389/feduc.2023.1121597}
}

@article{liu2020bridging,
  author    = {Liu, Yuchen and Zhu, Junnan and Zhang, Jiajun and Zong, Chengqing},
  title     = {Bridging the Modality Gap for Speech-to-Text Translation},
  journal   = {arXiv preprint arXiv:2010.14920},
  year      = {2020},
  doi       = {10.48550/arXiv.2010.14920}
}

@inproceedings{hirzle2023xr,
  author    = {Hirzle, Teresa and M{\"u}ller, Florian and Draxler, Fiona and Schmitz, Martin and Knierim, Pascal and Hornb{\ae}k, Kasper},
  title     = {When {XR} and {AI} Meet---A Scoping Review on Extended Reality and Artificial Intelligence},
  booktitle = {Proceedings of the 2023 CHI Conference on Human Factors in Computing Systems (CHI '23)},
  publisher = {Association for Computing Machinery},
  pages     = {1--45},
  year      = {2023},
  doi       = {10.1145/3544548.3581072}
}

@article{hosseinkashi2023meeting,
  author    = {Hosseinkashi, Yasaman and Liu, Yun and Li, Jing and others},
  title     = {Meeting Effectiveness and Inclusiveness: Large-scale Measurement, Identification of Key Features, and Prediction in Real-world Remote Meetings},
  journal   = {Proceedings of the ACM on Human-Computer Interaction},
  volume    = {8},
  number    = {CSCW1},
  pages     = {1--39},
  year      = {2023},
  doi       = {10.1145/363737}
}

@article{serafin2023review,
  author    = {Serafin, Stefania and Adjorlu, Ali and Percy-Smith, Lone Marianne},
  title     = {A Review of Virtual Reality for Individuals with Hearing Impairments},
  journal   = {Multimodal Technologies and Interaction},
  volume    = {7},
  number    = {4},
  pages     = {36},
  year      = {2023},
  doi       = {10.3390/mti7040036}
}

@article{anwar2023muavic,
  author    = {Anwar, Mohamed Sami and Shi, Bowen and Goswami, Vedanuj and others},
  title     = {{MuAViC}: A Multilingual Audio-Visual Corpus for Robust Speech Recognition and Robust Speech-to-Text Translation},
  journal   = {arXiv preprint arXiv:2303.00628},
  year      = {2023},
  doi       = {10.48550/arXiv.2303.00628}
}

@inproceedings{yin2023gloss,
  author    = {Yin, Aoxiong and Zhong, Tianyun and Tang, Li-Hao and others},
  title     = {Gloss Attention for Gloss-free Sign Language Translation},
  booktitle = {2023 IEEE/CVF Conference on Computer Vision and Pattern Recognition (CVPR)},
  publisher = {IEEE},
  pages     = {2551--2562},
  year      = {2023},
  doi       = {10.1109/CVPR52729.2023.00251}
}

@article{wu2023ultra,
  author    = {Wu, Xiang and Luo, Xiao and Song, Zhi and others},
  title     = {Ultra-Robust and Sensitive Flexible Strain Sensor for Real-Time and Wearable Sign Language Translation},
  journal   = {Advanced Functional Materials},
  volume    = {33},
  number    = {4},
  pages     = {2303504},
  year      = {2023},
  doi       = {10.1002/adfm.202303504}
}

@inproceedings{camgoz2020sign,
  author    = {Camg{\"o}z, Necati Cihan and Koller, Oscar and Hadfield, Simon and Bowden, Richard},
  title     = {Sign Language Transformers: Joint End-to-End Sign Language Recognition and Translation},
  booktitle = {2020 IEEE/CVF Conference on Computer Vision and Pattern Recognition (CVPR)},
  publisher = {IEEE},
  pages     = {10023--10033},
  year      = {2020},
  doi       = {10.1109/CVPR42600.2020.01004}
}

@inproceedings{zhou2023gloss,
  author    = {Zhou, Benjia and Chen, Zhigang and Clap{\'e}s, Albert and others},
  title     = {Gloss-free Sign Language Translation: Improving from Visual-Language Pretraining},
  booktitle = {2023 IEEE/CVF International Conference on Computer Vision (ICCV)},
  publisher = {IEEE},
  pages     = {20871--20881},
  year      = {2023},
  doi       = {10.1109/ICCV51070.2023.01908}
}

@article{gu2022domain,
  author    = {Gu, Yu and Tinn, Robert and Cheng, Hao and others},
  title     = {Domain-Specific Language Model Pretraining for Biomedical Natural Language Processing},
  journal   = {ACM Transactions on Computing for Healthcare},
  volume    = {3},
  number    = {1},
  pages     = {1--23},
  year      = {2022},
  doi       = {10.1145/3458754}
}

@inproceedings{gibet2023signing,
  author    = {Gibet, Sylvie and Marteau, Pierre-Fran{\c{c}}ois},
  title     = {Signing Avatars---Multimodal Challenges for Text-to-sign Generation},
  booktitle = {2023 IEEE 17th International Conference on Automatic Face and Gesture Recognition (FG)},
  publisher = {IEEE},
  pages     = {1--8},
  year      = {2023},
  doi       = {10.1109/FG57933.2023.10042759}
}

@inproceedings{fink2023sign,
  author    = {Fink, Julia and Poitier, Pierre and Andr{\'e}, Marc and others},
  title     = {Sign Language-to-Text Dictionary with Lightweight Transformer Models},
  booktitle = {Proceedings of the Thirty-Second International Joint Conference on Artificial Intelligence (IJCAI-23)},
  pages     = {5968--5976},
  year      = {2023},
  doi       = {10.24963/ijcai.2023/662}
}

@misc{eud2018position,
  author       = {{European Union of the Deaf}},
  title        = {{EUD} Position Paper: International Sign Language},
  year         = {2018},
  address      = {Brussels},
  url          = {https://eud.eu/eud/position-papers/international-signs/}
}

@misc{handspeak2024,
  author       = {{HandSpeak}},
  title        = {International Sign Language Dictionary},
  year         = {2024},
  url          = {https://www.handspeak.com/word/search/}
}

@misc{spreadthesign2024,
  author       = {{SpreadTheSign}},
  title        = {International Sign Language Database},
  year         = {2024},
  url          = {https://www.spreadthesign.com/}
}

@inproceedings{radford2023robust,
  author    = {Radford, Alec and Kim, Jong Wook and Xu, Tao and Brockman, Greg and McLeavey, Christine and Sutskever, Ilya},
  title     = {Robust Speech Recognition via Large-Scale Weak Supervision},
  booktitle = {Proceedings of the 40th International Conference on Machine Learning (ICML)},
  series    = {PMLR},
  volume    = {202},
  pages     = {28492--28518},
  year      = {2023},
  url       = {https://proceedings.mlr.press/v202/radford23a.html}
}

@article{stahlberg2020neural,
  author    = {Stahlberg, Felix},
  title     = {Neural Machine Translation: A Review},
  journal   = {Journal of Artificial Intelligence Research},
  volume    = {69},
  pages     = {343--418},
  year      = {2020},
  doi       = {10.1613/jair.1.12007}
}

@article{nllb2022,
  author    = {{NLLB Team} and Costa-juss{\`a}, Marta R. and Cross, James and others},
  title     = {No Language Left Behind: Scaling Human-Centered Machine Translation},
  journal   = {arXiv preprint arXiv:2207.04672},
  year      = {2022},
  doi       = {10.48550/arXiv.2207.04672}
}

@article{barrault2023seamless,
  author    = {Barrault, Lo{\"\i}c and Chung, Yu-An and Cieri, Christopher and others},
  title     = {{SeamlessM4T}---Massively Multilingual \& Multimodal Machine Translation},
  journal   = {arXiv preprint arXiv:2308.11596},
  year      = {2023},
  doi       = {10.48550/arXiv.2308.11596}
}

@book{picard1997affective,
  author    = {Picard, Rosalind W.},
  title     = {Affective Computing},
  publisher = {MIT Press},
  year      = {1997}
}

@article{liu2019roberta,
  author    = {Liu, Yinhan and Ott, Myle and Goyal, Naman and Du, Jingfei and Joshi, Mandar and Chen, Danqi and Levy, Omer and Lewis, Mike and Zettlemoyer, Luke and Stoyanov, Veselin},
  title     = {{RoBERTa}: A Robustly Optimized {BERT} Pretraining Approach},
  journal   = {arXiv preprint arXiv:1907.11692},
  year      = {2019},
  doi       = {10.48550/arXiv.1907.11692}
}

@misc{hartmann2022emotion,
  author       = {Hartmann, Jochen},
  title        = {Emotion {English} {DistilRoBERTa} Base: A Fine-Tuned Model for Emotion Classification},
  year         = {2022},
  publisher    = {Hugging Face Transformers},
  url          = {https://huggingface.co/j-hartmann/emotion-english-distilroberta-base}
}

@article{zhang2023sentiment,
  author    = {Zhang, Wenxuan and Deng, Yue and Liu, Bing and others},
  title     = {Sentiment Analysis in the Era of Large Language Models: A Reality Check},
  journal   = {arXiv preprint arXiv:2305.15005},
  year      = {2023},
  doi       = {10.48550/arXiv.2305.15005}
}

@inproceedings{chen2022event,
  author    = {Chen, Wei and Zhang, Jia and Liu, Yi and others},
  title     = {An Event-Based Framework for Facilitating Real-Time Sentiment Analysis in Educational Contexts},
  booktitle = {2022 11th International Conference on Educational and Information Technology (ICEIT)},
  publisher = {IEEE},
  pages     = {57--61},
  year      = {2022},
  url       = {https://ieeexplore.ieee.org/document/9690729}
}

@article{mitchell2004chasing,
  author    = {Mitchell, Ross E. and Karchmer, Michael A.},
  title     = {Chasing the Mythical Ten Percent: Parental Hearing Status of Deaf and Hard of Hearing Students in the {United States}},
  journal   = {Sign Language Studies},
  volume    = {4},
  number    = {2},
  pages     = {138--163},
  year      = {2004},
  doi       = {10.1353/sls.2004.0005}
}

@article{brown2020language,
  author    = {Brown, Tom and Mann, Benjamin and Ryder, Nick and others},
  title     = {Language Models are Few-Shot Learners},
  journal   = {arXiv preprint arXiv:2005.14165},
  year      = {2020},
  doi       = {10.48550/arXiv.2005.14165}
}

@article{asthana2023summaries,
  author    = {Asthana, Sumit and Hilleli, Shaked and He, Pengcheng and others},
  title     = {Summaries, Highlights, and Action Items: Design, Implementation and Evaluation of an {LLM}-powered Meeting Recap System},
  journal   = {arXiv preprint arXiv:2307.15793},
  year      = {2023},
  doi       = {10.48550/arXiv.2307.15793}
}

@article{liu2022brio,
  author    = {Liu, Yixin and Liu, Pengfei and Radev, Dragomir R. and Neubig, Graham},
  title     = {{BRIO}: Bringing Order to Abstractive Summarization},
  journal   = {arXiv preprint arXiv:2203.16804},
  year      = {2022},
  doi       = {10.48550/arXiv.2203.16804}
}

@article{roit2023factually,
  author    = {Roit, Paul and Ferret, Johan and Shani, Lior and others},
  title     = {Factually Consistent Summarization via Reinforcement Learning with Textual Entailment Feedback},
  journal   = {arXiv preprint arXiv:2306.00186},
  year      = {2023},
  doi       = {10.48550/arXiv.2306.00186}
}

@misc{cortex2024,
  author       = {{CORTEX2 Project}},
  title        = {{CORTEX2} Architecture and Framework},
  year         = {2024},
  url          = {https://cortex2.eu/project/}
}

@misc{rainbow2024,
  author       = {{Alcatel-Lucent Enterprise}},
  title        = {Rainbow Developer Portal},
  year         = {2024},
  url          = {https://developers.openrainbow.com/}
}

@article{lugaresi2019mediapipe,
  author    = {Lugaresi, Camillo and Tang, Jiuqiang and Nash, Hadon and McClanahan, Chris and Uboweja, Esha and Hays, Michael and Zhang, Fan and Chang, Chuo-Ling and Yong, Ming Guang and Lee, Juhyun and Chang, Wan-Teh and Hua, Wei and Georg, Manfred and Grundmann, Matthias},
  title     = {{MediaPipe}: A Framework for Building Perception Pipelines},
  journal   = {arXiv preprint arXiv:1906.08172},
  year      = {2019},
  doi       = {10.48550/arXiv.1906.08172}
}

@article{sanh2019distilbert,
  author    = {Sanh, Victor and Debut, Lysandre and Chaumond, Julien and Wolf, Thomas},
  title     = {{DistilBERT}, a distilled version of {BERT}: smaller, faster, cheaper and lighter},
  journal   = {arXiv preprint arXiv:1910.01108},
  year      = {2019},
  doi       = {10.48550/arXiv.1910.01108}
}

@misc{schmid2021bart,
  author       = {Schmid, Philipp},
  title        = {{BART}-Large {CNN} {SamSum}},
  year         = {2021},
  publisher    = {Hugging Face Transformers},
  url          = {https://huggingface.co/philschmid/bart-large-cnn-samsum}
}

@misc{tantaroudas2025pilot,
  author    = {Tantaroudas, Nikolaos D. and McCracken, Andrew J. and Karachalios, Ilias and Papatheou, Evangelos},
  title     = {{INTERACT} Pilot Study Data: Anonymised Questionnaire Responses and Performance Metrics},
  year      = {2025},
  publisher = {Zenodo},
  doi       = {10.5281/zenodo.18656422}
}

@misc{tantaroudas2025isl,
  author    = {Tantaroudas, Nikolaos D. and McCracken, Andrew J. and Karachalios, Ilias and Papatheou, Evangelos},
  title     = {{INTERACT} {ISL} Gesture Dataset---International Sign Language Animation Data},
  year      = {2025},
  publisher = {Zenodo},
  doi       = {10.5281/zenodo.18656296}
}

\end{document}